\documentstyle[12pt]{article}

\makeatletter
\ifcase\@ptsize
  \font\tenmsy=msbm10
  \font\sevenmsy=msbm7
  \font\fivemsy=msbm5
  \font\tenmsx=msam10
  \font\sevenmsx=msam7
  \font\fivemsx=msbm5
\or
  \font\tenmsy=msbm10 scaled \magstephalf
  \font\sevenmsy=msbm8
  \font\fivemsy=msbm6
  \font\tenmsx=msam10 scaled \magstephalf
  \font\sevenmsx=msam8
  \font\fivemsx=msam6
\or
  \font\tenmsy=msbm10 scaled \magstep1
  \font\sevenmsy=msbm8
  \font\fivemsy=msbm6
  \font\tenmsx=msam10 scaled \magstep1
  \font\sevenmsx=msam8
  \font\fivemsx=msam6
\fi
\newfam\msyfam
\textfont\msyfam=\tenmsy
\scriptfont\msyfam=\sevenmsy
\scriptscriptfont\msyfam=\fivemsy
\newfam\msxfam
\textfont\msxfam=\tenmsx
\scriptfont\msxfam=\sevenmsx
\scriptscriptfont\msxfam=\fivemsx
\def\Bbb{\ifmmode\let\next\Bbb@\else
\def\next{\errmessage{Use \string\Bbb\space only in math mode}}\fi\next}
\def\Bbb@#1{{\Bbb@@{#1}}}
\def\Bbb@@#1{\fam\msyfam#1}
\def\Aaa{\ifmmode\let\next\Aaa@\else
\def\next{\errmessage{Use \string\Aaa\space only in math mode}}\fi\next}
\def\Aaa@#1{{\Aaa@@{#1}}}
\def\Aaa@@#1{\fam\msxfam#1}
\newfam\euffam
\font\sixeuf=eufm6
\font\eighteuf=eufm8
\font\twelveeuf=eufm10 scaled\magstep1
\textfont\euffam=\twelveeuf
\scriptfont\euffam=\eighteuf
\scriptscriptfont\euffam=\sixeuf

\makeatother
\newcommand{\BC}{{\Bbb{C}}}

\newcommand{\BI}{{\Bbb{I}}}

\newcommand{\BP}{{\Bbb{P}}}

\newcommand{\BR}{{\Bbb{R}}}
\newcommand{\BZ}{{\Bbb{Z}}}

\def\nn{\noindent}

\def\vec#1{{\rm\bf#1}}

\def\vev#1{\langle{#1}\rangle}

\def\Vev#1{\left\langle{#1}\right\rangle}

\def\Vvev#1{\left\langle\kern-5pt\left\langle{%
  #1}\right\rangle\kern-5pt\right\rangle}
\def\Vvevs#1{\left\langle\kern-4pt\left\langle{%
  #1}\right\rangle\kern-4pt\right\rangle}
\def\Vvevss#1{\left\langle\kern-3pt\left\langle{%
  #1}\right\rangle\kern-3pt\right\rangle}

\def\be{\begin{equation}}
\def\ee{\end{equation}}
\def\ba{\begin{array}}
\def\ea{\end{array}}
\def\bea{\begin{eqnarray}}
\def\eea{\end{eqnarray}}
\def\bean{\begin{eqnarray*}}
\def\eean{\end{eqnarray*}}
\def\bl{\begin{list}{}{}}

\def\ts{\textstyle}

\newcommand{\reseteqn}{\setcounter{equation}{0}\setcounter{subsection}{0}
\setcounter{subsubsection}{0}}
\newcommand{\mysection}{\reseteqn\section}

\if@twoside
\else
\fi
\begin{document}
  \pagestyle{empty}
  \begin{raggedleft}
KCL-MTH-98-36\\
hep-th/9808169\\
August 1998\\
  \end{raggedleft}\par
  \vfill
  {\Large\sc
  \begin{center}
Logarithmic Conformal Field Theory\\ \&\\ Seiberg-Witten Models
  \end{center}
  }\par
  \vfill
  \begin{center}
{\sc Mic$\hbar$ael~A.I.~Flohr\footnote[1]{
  \begin{tabular}[t]{rl}
    homepage: & {\tt http://www.mth.kcl.ac.uk/\~{}flohr/}\\
    email:    & {\tt flohr@mth.kcl.ac.uk}
  \end{tabular}
}}\\
$\phantom{X}$\\
{\em Department of Mathematics\\
King's College London\\
The Strand\\
London WC2R 2LS, United Kingdom}
  \end{center}\par
  \vfill
  \begin{abstract}
\noindent The periods of arbitrary abelian forms on hyperelliptic
Riemann surfaces, in particular the periods of the meromorphic
Seiberg-Witten differential $\lambda_{{\rm SW}}$, are shown to be
in one-to-one correspondence with the conformal blocks of correlation
functions of the rational logarithmic conformal field theory with
central charge $c=c_{2,1}=-2$. The fields of this theory precisely
simulate the branched double covering picture of a hyperelliptic curve,
such that generic periods can be expressed in terms of certain generalised
hypergeometric functions, namely the Lauricella functions of type $F_D$.
  \end{abstract}
  \vfill
  \newpage
%
%
  \setcounter{page}{1}
  \pagestyle{plain}
  \mysection{{\sc Introduction}}
\vspace{-7pt}
\nn In a seminal work \cite{SW94}, Seiberg and Witten found the exact
low-energy effective action of four-dimensional $N$=2 supersymmetric
$SU(2)$ Yang-Mills theory. Soon, this was generalised to general
simple gauge groups \cite{SWetc}. At the heart of the exact solution lies a
certain Riemann surface, in the case of a simple, simply-laced gauge
group a hyperelliptic one, which constitutes the moduli space of
the Yang-Mills theory. All information, in particular the scalar
modes and the prepotential, are encoded in this hyperelliptic curve
and a special meromorphic differential form associated to it, the
so-called Seiberg-Witten differential $\lambda_{{\rm SW}}$.
The task of exactly solving the low-energy effective field theory
is then reduced to essentially computing the periods of
$\lambda_{{\rm SW}}$.

In this paper, we will achieve the computation of the Seiberg-Witten
periods in a new way, expressing them in terms of conformal blocks
of a very special conformal field theory (CFT) with central charge
$c=-2$. This theory belongs to a rather new class of CFTs, which has
been studied in some detail only recently \cite{LCFTgen},
the so-called logarithmic
conformal field theories (LCFTs).
First encountered and shown to be consistent in \cite{Gur93},
they are not just a peculiarity but merely a generalisation of
ordinary two-dimensional CFTs with broad and growing applications
\cite{LCFTapp}.
As is particularly true for Seiberg-Witten models, logarithmic divergences
are sometimes quite physical, and so there is an increasing interest in
these logarithmic conformal field theories. The relevance of LCFT in the
Seiberg-Witten context has first been observed in \cite{Cappelli}.

Furthermore, this application illuminates the geometry behind logarithmic
CFT. It is well known that vertex operators of worldsheet CFTs in string
theory describe the equivalent of Feynman graphs with outer legs by
simulating their effect on a Riemann surface as punctures. Now, in the new
setting of moduli spaces of low-energy effective field theories, {\em pairs
of\/} vertex operators describe the insertion of additional handles to a
Riemann surface, simulating the resulting branch cuts. So, in much the same
way as a smooth but infinitely long stretched tube attached to an otherwise
closed worldsheet, standing for an external state, is replaced by a
puncture with an appropriate vertex operator, so a smooth additional
handle, standing for an intersecting 4-brane on the 5-brane worldvolume
in the type IIA picture of low-energy effective field theories, is
replaced by branch cuts with appropriate vertex operators at its endpoints.
Hence, operator product expansions (OPEs) of such vertex operators
simulating branch points, poles etc.\ on the curve
represented as a branched covering
$Z:\Sigma\rightarrow\BC\BP^1$ provide an intuitive way of
understanding what happens when, for instance, intersecting 4-branes
run into each other or shrink to zero size (implying the same for
the branch cuts).

This letter is organised as follows: In section II we briefly discuss
the hyperelliptic curves and the Seiberg-Witten differential in the form
relevant to our approach. Section III recapitulates the construction of
1-differentials on hyperelliptic curves in terms of vertex operators,
emphasising why this leads to a logarithmic CFT. Then we have all material
at hand to actually compute the Seiberg-Witten periods in terms of
conformal blocks in section IV, also expressing them in terms of certain
special functions. We conclude this last section with a brief discussion
and outlook. 
This letter is a short version, loosely based on several talks held at Durham,
King's College London, Oxford, and SISSA Trieste, of a much more 
detailed and rigorous work to appear soon \cite{myself}.

%
%
  \mysection{{\sc Seiberg-Witten Solutions of
  Supersymmetric\\ Four-Dimensional Yang-Mills Theories}}

\nn In a much celebrated work \cite{SW94}, Seiberg and Witten found an
exact solution
to $N$=$2$ supersymmetric four-dimensional Yang-Mill theory with gauge group
$SU(2)$. This paper initiated intensive research \cite{SWetc}
leading to a vast set of exactly soluble Yang-Mills theories
in various dimensions and with various degrees of supersymmetry. Of
particular interest for these solutions is the understanding of the moduli
space of vacua, which in many cases turns out to be a hyperelliptic
Riemann surface.

The BPS spectrum of such a model is entirely determined by the periods of
a special meromorphic 1-differential on this Riemann surface, the famous
Seiberg-Witten differential $\lambda_{{\rm SW}}$, which yields the
scalar modes. Let $\alpha_i, \beta^j$ denote a canonical basis of the
homology of the Riemann surface, $\alpha_i\cap\beta^j=\delta_i^{\phantom{i}j}$,
then the scalar modes are simply given as $a_i = \oint_{\alpha_i}
\lambda_{{\rm SW}}$, $a_D^j=\oint_{\beta^j}\lambda_{{\rm SW}}$.
These scalar modes carry electric and magnetic charges respectively,
and the mass of a BPS state with charges $(\vec{q},\vec{g})$ is then
given as $m_{(\vec{q},\vec{g})} \sim |q^i a_i + g_j a_D^j|$, momentarily
neglecting possible residue terms in case of the presence of hypermultiplets.

A general hyperelliptic
Riemann surface can be described in terms of two variables $w,Z$ in the
polynomial form
  \be\label{eq:surf}
    w^2 + 2A(Z)w + B(Z) = 0
  \ee
with $A(Z),B(Z)\in\BC[Z]$. After
a simple coordinate transformation in $y=w+A(Z)$, this takes on the
more familiar form $y^2 = A(Z)^2 - B(Z)$. But we might also write
the hyperelliptic curve in terms of a rational map if we divide the
defining equation (\ref{eq:surf}) by $A(Z)^2$ and put $\tilde{w}=w/A(Z)+1$
to arrive at the representation
  \be\label{eq:ratmap}
    (1-\tilde{w})(1+\tilde{w}) = \frac{B(Z)}{A(Z)^2}\,.
  \ee
This form is very appropriate in the frame of Seiberg-Witten models, since
the Seiberg-Witten differential can be read off directly: The rational map
$R(Z)=B(Z)/A(Z)^2$ is singular at the zeroes of $B(Z)$ and $A(Z)$, and is
degenerate whenever its Wronskian $W(R) \equiv W(A(Z)^2,B(Z)) =
(\partial_ZA(Z)^2)B(Z) - A(Z)^2(\partial_ZB(Z))$ vanishes. This is precisely
the information encoded in $\lambda_{{\rm SW}}$ which for arbitrary
hyperelliptic curves, given by a rational map $R(Z)=B(Z)/A(Z)^2$, can be
expressed as
  \be\label{eq:lamSW}
    \lambda_{{\rm SW}} =
    \frac{Z}{2\pi i}\,{\rm d}(\log\frac{1-\tilde{w}}{1+\tilde{w}}) =
    \frac{1}{2\pi i}{\rm d}(\log R(Z))\frac{Z}{\tilde{w}}
    = \frac{1}{2\pi i}
    \frac{W(A(Z)^2,B(Z))}{A(Z)B(Z)}\frac{Z\,{\rm d}Z}{y}\,.
  \ee
Note that the fact that the denominator polynomial is a square guarantees
the curve to be hyperelliptic. It is this local form of the Seiberg-Witten
differential which serves as a metric ${\rm d}s^2 = |\lambda_{{\rm SW}}|^2$
on the Riemann surface, and it is this local form which arises as the
tension of self-dual strings coming from 3-branes in type II string theory
compactifications on Calabi-Yau threefolds.\footnote[1]{
This derivation of the Seiberg-Witten differential is equivalent
to the one from integrable Toda systems with spectral curve $z+1/z+r(t) =
z+1/z+2A(t)/\sqrt{B(t)}=0$, where $\lambda_{{\rm SW}}=t\,{\rm d}(\log\,z)$
is nothing other than the Hamilton-Jacobi function of the underlying integrable
hierarchy. However, the price paid for this very simple form of $\lambda_{{\rm
SW}}$ is that $r(t)$ is now only a fractional rational map.}

Let us, for the sake
of simplicity, concentrate on $N$=$2$ $SU(N_c)$ Yang-Mills theory with
$N_f$ massive hypermultiplets. Then, the hyperelliptic curve $y^2=A(x)^2-B(x)$
takes the form
  \be
    y^2 = \left(x^{N_c} - \sum_{k=2}^{N_c}s_k x^{N_c-k}\right)^2
          -\Lambda^{2N_c-N_f}\prod_{i=1}^{N_f}(x-m_i)
        = \prod_{j=1}^{2N_c}(x-e_j)\,,
  \ee
where we have absorbed any dependency of $A(x)=\prod_{k=1}^{N_c}(x
-\tilde{a}_k)$ on
the $m_i$, which is the case for $N_f > N_c$, in a redefinition of the
classical expectation values $\tilde{a}_k$
or $s_k$ respectively. Then, the Seiberg-Witten differential takes the
general form
  \be
    \lambda_{{\rm SW}}(SU(N_c)) = \frac{1}{2\pi i}
    \frac{\prod_{l=0}^{N_c+N_f-1}(x-z_l)}{
    \prod_{j=1}^{2N_c}\sqrt{x-e_j}\prod_{i=1}^{N_f}(x-m_i)}\,{\rm d}x\,,
  \ee
where the $z_l$ denote the zeroes of $2A(x)'B(x)-A(x)B(x)'$, and $z_0=0$.
As a result, the total order of the general Seiberg-Witten form
(\ref{eq:lamSW}) vanishes,
$(1 + N_c+N_f-1)\cdot(1) + (2N_c)\cdot(-\frac12) + (N_f)\cdot(-1) = 0$
implying that $\lambda_{{\rm SW}}$ has a double pole at infinity,
which will be important later. We note that the periods of the
Seiberg-Witten form are hence contour integrals with paths encircling
pairs $(e_i,e_j)$ and with an integral kernel of the form
  \be
    \lambda_{{\rm SW}} \sim \prod_i(x-x_i)^{r_i}\,,\ \ \ \
    \sum_i r_i = 0\,,\ \ \ \ 
    r_i\in\{{\ts 0,\pm\frac12,\pm 1}\}\,,
  \ee
where the branch points $e_i$ are a subset of the singular points
$x_i$ of the integral kernel.

%
%
  \mysection{{\sc The $c=-2$ Logarithmic CFT and 1-Differentials}}

\nn The idea to represent general $j$-differentials ($j\in\BZ/2$ due
to locality) in
terms of fields of a CFT is actually not new. We will follow here the
approach put forward by Knizhnik \cite{Kniz},
restricted to the case of interest,
$j=1$ and hyperelliptic curves, i.e.\ all branch points have
ramification number two. As we will demonstrate,
this CFT approach to the theory of Riemann surfaces naturally leads
to a logarithmic CFT. This is a crucial fact which can only be
appreciated now, after the advent of LCFT.

In the case of hyperelliptic curves, $j$-differentials are constructed
by two pairs of anticommuting fields $\phi^{(j),\ell},\phi^{(1-j),\ell}$
of spin $j,1-j$ respectively, one pair for each sheet of the Riemann
surface $\Sigma$ represented as a branched covering of $\BC\BP^1$,
where the sheets are enumerated by $\ell=0,1$.
We will denote the covering map by $Z$. The point is that such fields
behave as differentials of weight $j$ under conformal transformations,
  \be
    \phi^{(j),\ell}(Z',\bar Z')\left(\frac{{\rm d}Z'}{{\rm d}Z}\right)^j =
    \phi^{(j),\ell}(Z,\bar Z)\,.
  \ee
We assume that the operator product expansion (OPE) be normalised as
  \be\label{eq:openorm}
    \phi^{(j),\ell}(Z')\phi^{(1-j),\ell}(Z)\simeq\BI\,(Z'-Z)^{-1} +
    {\rm regular\ terms}
  \ee
with $\BI$ denoting the identity operator.
On each sheet, we have an action
  \be
    S^{(\ell)} = \int\,\phi^{(j),\ell}\,\bar{\partial}\phi^{(1-j),\ell}\,
    {\rm d}^2Z = \int\,\phi^{(1),\ell}\,\bar{\partial}\phi^{(0),\ell}\,,
  \ee
where integration runs over the Riemann surface $\Sigma$,
and a stress energy tensor which takes the form
  \be
    T^{(\ell)} = -j\phi^{(j),\ell}\,\partial\phi^{(1-j),\ell}
               + (j-1)\phi^{(1-j),\ell}\,\partial\phi^{(j),\ell}
               = -\phi^{(1),\ell}\,\partial\phi^{(0),\ell}
  \ee
giving rise to a central extension $c=c_j\equiv-2(6j^2-6j+1)$, i.e.\ in our
case $c=c_1 = -2$.

Let now a hyperelliptic curve of genus $g$ be given as
$y^2=\prod_{k=1}^{2g+2}(Z-e_k)$ such that infinity would not be a branch
point. At each branch point $e_k$, we can
locally invert this to $Z(y) \sim e_k + y^2$ such that we have in the
vicinity of $e_k$ that $y(Z) \sim (z-e_k)^{1/2}$. Let us denote the
operation of moving a point around $e_k$ by $\hat{\pi}_{e_k}$. This
operation acts on the $j$-differentials with the following boundary
conditions:
  \be
    \hat{\pi}_{e_k}\phi^{(j),\ell}(Z)=
    (-)^{2j}\phi^{(j),\ell+1\,{\rm mod}\,2}(Z)
  \ee
in the vicinity of $e_k$.
Since all branch points have the same ramification number two, i.e.\ the
$\BZ_2$ symmetry of $\Sigma$ is global, we can diagonalize $\hat{\pi}$
globally by choosing a new basis via a discrete Fourier
transform,
  \be
    \phi^{(j)}_k = \phi^{(j),0} + (-)^{j-k}\phi^{(j),1}
    \,,
  \ee
with $k=0,1$,
such that $\hat{\pi}_{a}\phi^{(j)}_k = (-)^{k-j}\phi^{(j)}_k$
for $a$ any branch point.
We can now define chiral currents $J_k =
\mbox{:$\phi^{(j)}_k\phi^{(1-j)}_k$:}$,
$\bar{\partial}J_k = 0$, which are single valued functions near $a$.
It follows then that a branch point $a$ carries charges
$q_k = \frac12(j-k) = \frac12(1-k)$ with respect to these currents.

In order to do explicit calculations it is helpful to
bosonize with the help of two analytic scalar
fields $\varphi_k$, $k=0,1$, normalised in the usual way
$\vev{\varphi_k(z)\varphi_l(z')}=-\delta_{kl}\log(z-z')$. It is then easy
to see that we have $\phi^{(j)}_k = \mbox{:$\exp(-i\varphi_k)$:}$,
$\phi^{(1-j)}_{1-k} = \mbox{:$\exp(+i\varphi_k)$:}$,
$J_k = i\partial\varphi_k$, and $T_k = \frac12 \mbox{:$J_kJ_k$:}
+\frac12 \partial J_k$. Hence, we have
a Coulomb gas CFT with background charge $2\alpha_0=1$. In general we
define vertex operators with charge $\vec{q}=(q_0,q_1)$ as
$V_{\vec{q}}(a) =
\mbox{:$\exp(i\,\vec{q}\!\cdot\!\mbox{{\boldmath $\varphi$}}(a))$:}$
which have conformal scaling dimensions
$h(\vec{q}) = h_0 + h_1$ with $h_k = \frac12(q_k^2 - q_k^{})$.
Note that branch points are trivial objects in the $k=1$ sector such that
it suffices to only consider the $k=0$ sector from now on.

If one now tries to proceed in the usual manner, one seems to run into
a crucial obstacle. It is well known that correlators in free field
realization of CFT are only non-zero, if they satisfy the charge neutrality
condition. For example, the only non-vanishing two-point functions are
$\vev{V_{2\alpha_0-q}(z)V_q(z')} = A(z-z')^{-2h(q)}$, where $A$ usually can
be chosen arbitrarily by normalisation of the fields. However, a careful
analysis shows \cite{KoLe97} that the vertex operator which represents
a branch point does not have a conjugate field as we expect it. The
charge of a branch point is simply $q=\alpha_0=1/2$ such that
$2\alpha_0-q=q$,
i.e.\ the branch point vertex operator appears to be self-conjugate.
However, this is not true, $\vev{V_{1/2}(z)V_{1/2}(z')}=0$.
It turns out that the correct partner of this field is
$\Lambda_{1/2} = \left.\partial_q V_q\right|_{q=\alpha_0}=
i\varphi V_{1/2}$, such that
$\vev{\Lambda_{1/2}(z)V_{1/2}(z')}=B(z-z')^{1/4}$ and
$\vev{\Lambda_{1/2}(z)\Lambda_{1/2}(z')}=
(C-2B\log(z-z'))(z-z')^{1/4}$. The constants $A,B,C$ are now
no longer entirely free. $SL(2,\BC)$ invariance of the two-point
functions requires that $A=0$, $B=\vev{2i\varphi V_{2\alpha_0}}=1$, $C=0$.
Although this field $\Lambda_{1/2}$ is a proper primary
field with respect to the stress energy tensor, it will cause
logarithmic terms in the OPE with other primary fields. It will also
give rise to other fields of this form, $\Lambda_q = (\partial_q h(q))^{-1}
\partial_q V_q = \frac{i}{q-\alpha_0}\varphi V_q$ which are the logarithmic
partners to the primary fields $V_{1-q}$. Note that the latter definition
of $\Lambda_q$ is only valid for $q\neq\alpha_0=\frac12$.
A special feature of this CFT is that the conformal Ward identities
force us to put $\vev{V_1} = \vev{V_0} = \vev{\BI} = 0$, while
$\vev{\Lambda_1} = 1$. This might seem strange but can be seen to be
quite natural in the original definition of this CFT (before
bosonization), or even better in a realization of it by a pair of
anticommuting scalar fields with manifest $SL(2,\BC)$ invariance, where
the path integrals vanish unless zero modes are inserted \cite{GFN98}.
In fact, the naive definition $\det \bar{\partial}_{(j)} = \int
{\cal D}\phi^{(j),\ell}{\cal D}\phi^{(1-j),\ell}\exp(S^{(\ell)})$ vanishes,
due to $n_j-n_{1-j}=(2j-1)(g-1)$ zero modes of $\bar{\partial}$-holomorphic
$j$- and $(1-j)$-differentials on a genus $g$ Riemann surface.

To summarise, the $c=-2$ CFT of 1-differentials inevitably becomes
logarithmic when we add to its field content the vertex operator $V_{1/2}$
which represents branch points of a hyperelliptic curve. The reason is
that adding this vertex operator yields vanishing or trivial correlation
functions unless we also introduce its proper conjugate field $\Lambda_{1/2}$
which helps to cancel off the $n_{1-j}=1$ scalar zero mode.
For example, only such 4-point correlators are non-zero which
contain one and only one scalar zero mode, i.e.\ one and only one of
the fields $\Lambda_{1/2}$. More generally, reducing an arbitrary correlation
function with vertex operators $V_q$ and logarithmic partners
$\Lambda_q$ ultimately will result in picking out only such nested OPEs,
which lead to the only non-vanishing one-point functions $\vev{\Lambda_q}$.
For example, the logarithmic partner of the identity, $\Lambda_1$, has
the OPE $\Lambda_1(z)\Lambda_1(z') = \BI -2\log(z-z')\Lambda_1(z') + \ldots$
without any term of the form :\mbox{$\varphi\varphi\exp(2i\varphi)$}: which
would lead to multiple logarithms. Hence, $\vev{\Lambda_1(z)\Lambda_1(z')}
= -2\log(z-z')\vev{\Lambda_1} = -2\log(z-z')$.

We will adopt the following conventions: First, from the above follows
that we can replace the operator for a branch point by $\mu(a) =
V_{1/2}(a) + \Lambda_{1/2}(a)$. Next, we will introduce the reduced
correlators
  \be\label{eq:vvev}
  \Vvev{\prod_i \Phi_{q_i}(z_i)} \equiv
  \prod_{k<l}(z_k-z_l)^{-q_kq_l}\Vev{\prod_i \Phi_{q_i}(z_i)}
  \ee
where the canonical free part has been divided off, $\Phi=V,\Lambda$.
The reduced correlator is thus equal to the screening charge integrals
still necessary to ensure charge neutrality.
Under conformal transformations $z\mapsto M(z)$, a correlator transforms
with weights $\left(\left.\partial_z M(z)\right|_{z=z_i}\right)^{h(q_i)}$
for each field $\Phi_{q_i}(z_i)$. For the reduced correlators, the exponent
simply has to be replaced by $-q_i/2$.

We are now in the position to express an arbitrary abelian differential
on the hyperelliptic curve $\Sigma:y^2=\prod_{k=1}^{2g+2}(Z-e_k)=
\prod_{k=1}^{g+1}(Z-e_k^-)(Z-e_k^+)$ in
terms of fields of the $c=-2$ LCFT. In fact, we have with the above
notations
  \be
    \omega = \frac{\prod_{i=1}^{M}(Z-z_i)}{
    \prod_{k=1}^{2g+2}\sqrt{Z-e_k}\prod_{j=1}^{N}(Z-p_j)}\,{\rm d}Z =
    \prod_{i=1}^{M}V_{-1}(z_i)\prod_{k=1}^{2g+2}\mu(e_k)
    \prod_{j=1}^{N}V_{1}(p_j)\,\phi^{(1)}_0(Z)\,.
  \ee
In case that one of the zeroes coincides with a branch point,
we replace according to the OPE $\,\lim_{z_i\rightarrow e_k}(z_i-e_k)^{1/2}
V_{-1}(z_i)\mu(e_k) = V_{-1/2}(e_k)+\Lambda_{-1/2}(e_k)\equiv\sigma(e_k)$.
It is then clear that a contour
integral along a closed path $\gamma$ defines a conformal block
  \be\label{eq:periods}
    \oint_{\gamma}\omega =
    \Vvev{V_Q(\infty)\prod_{i=1}^{M}V_{-1}(z_i)
    \prod_{k=1}^{2g+2}\mu(e_k)
    \prod_{j=1}^{N}V_{1}(p_j)}_{(\gamma)}\,,
  \ee
where $Q=2-\sum q_i=1+M-N-g$ is the charge of a pole
at infinity such that charge neutrality is ensured by insertion of only
one screening charge $Q_-=\oint J_-$ with $J_-\equiv\phi^{(1)}_0$ being
the 1-differential (note that $2\alpha_0=1$ and that $\phi^{(1)}_0\sim
V_{-1}$ changes the charge by $-1$). We now choose (part of)\footnote[1]{
The integral kernel $\omega$ has further singular points $z_i,p_j$.
Although the former can be multiplied out to yield a sum of smaller
integral kernels, and although the latter simply contribute residual
terms, we can treat them on equal footing with the branch points $e_k$ in
the CFT picture by analytic continuation of correlation functions
with $q_i\not\in\BZ/2$ to these particular values. Of course, this
enlarges the number of possible contours and hence possible conformal
blocks.} the basis of
conformal blocks to coincide with the canonical homology basis of cycles,
i.e.\ $\gamma\in\{\alpha_i,\beta^i\}_{1\leq i\leq g}$ which can be
choosen as $\alpha_i = C_{(e_i^-,e_i^+)}$, $\beta^i = C_{(e_i^+,e_{g+1}^-)}$.
Here, $C_{(a,b)}$ denotes a closed path encircling $a,b$.

%
%
  \mysection{{\sc Periods of the Seiberg-Witten Differential}}

\nn Let us start with a warm up by calculating the periods of the
only holomorphic one-form for the torus, i.e.\ $g=1$ and the gauge group
is $SU(2)$. The torus in question is given by $y^2 = (x^2-u)^2 - \Lambda^4$
with the four branch points $e_1=\sqrt{u-\Lambda^2}$,
$e_2=-\sqrt{u+\Lambda^2}$, $e_3=-\sqrt{u-\Lambda^2}$, $e_4=\sqrt{u+\Lambda^2}$.
The standard periods of the only holomorphic form, ${\rm d}x/y$,
are easily computed (where the normalization has been fixed to be in
accordance with the asymptotic behavior of $a$ and $a_D$ in the weak
coupling region):
  \bea
    \pi_1 = \frac{\partial a}{\partial u}
          &=& \frac{\sqrt{2}}{2\pi}\int_{e_2}^{e_3}\frac{{\rm d}x}{y}
          \ =\ \frac{\sqrt{2}}{2\pi}
              \Vvevss{\mu(e_1)\mu(e_2)\mu(e_3)\mu(e_4)}_{(e_2,e_3)}
              \nonumber\\
          &=& \frac{\sqrt{2}}{2\pi}
              (e_3-e_2)^{-\frac{1}{2}}(e_4-e_1)^{-\frac{1}{2}}
              \Vvevss{\mu(\infty)\mu(1)\mu(0)\mu(M(e_4))}_{(0,1)}
              \nonumber\\
          &=& \frac{\sqrt{2}}{2}
              (e_2-e_1)^{-\frac{1}{2}}(e_4-e_3)^{-\frac{1}{2}}
              {}_2F_1({\textstyle\frac12,\frac12};1;\xi)\,,
  \eea
where $\xi=1/M(e_4)=\frac{(e_1-e_4)(e_3-e_2)}{(e_2-e_1)(e_4-e_3)}$ is the
inverse crossing ratio,
$\xi=(u-\sqrt{u^2-\Lambda^4})/(u+\sqrt{u^2-\Lambda^4})$.
The other period is obtained in complete analogy by exchanging $e_2$ with
$e_1$, yielding
  \be
    \pi_2 = \frac{\partial a_D}{\partial u} =
        \frac{\sqrt{2}}{2\pi}\int_{e_1}^{e_3}\frac{{\rm d}x}{y} =
        \frac{\sqrt{2}}{2}(e_1-e_2)^{-\frac{1}{2}}(e_4-e_3)^{-\frac{1}{2}}
        {}_2F_1({\textstyle\frac12,\frac12};1;1-\xi)\,.
  \ee
Here and in the following, (generalized) hypergeometric functions with
arguments such as $1-\xi$ are understood as expansions around $1-\xi$ and
should be analytically continued to a region around $\xi$. This will result in
the desired logarithmic divergencies. For example, using the usual
Frobenius process, we find (the factor $\pi=\Gamma(\frac12)^2$ stems from
the formula for analytic continuation of hypergeometric functions)
  \bea
    \pi\,{}_2F_1({\textstyle\frac12,\frac12;1;1-\xi}) &=&
    {}_2F_1({\textstyle\frac12,\frac12;1;\xi})\log(\xi) +
    \sum_{n=0}^{\infty}\left.\left(\frac{\partial}{\partial\varepsilon}
    \frac{(\frac12+\varepsilon)_n(\frac12+\varepsilon)_n}
    {(1+\varepsilon)_n(1+\varepsilon)_n}\right)\right|_{\varepsilon=0}\xi^n
    \\
    &=& {}_2F_1({\textstyle\frac12,\frac12;1;\xi})\log(\xi) +
    \left.\partial_{\varepsilon}\,
    {}_3F_2({\textstyle 1,\frac12+\varepsilon,\frac12+\varepsilon;
    1+\varepsilon,1+\varepsilon;\xi})\right|_{\varepsilon=0}\nonumber\,.
  \eea
These results are, of course, well known. Less known might be the
fact that for the case without hyper-multiplets, $N_f=0$, we can express
the periods of the Seiberg-Witten form by the Lauricella function
$F_D^{(3)}$. In fact,
  \bea
    a(u) &=& \frac{\sqrt{2}}{2\pi}\int_{e_2}^{e_3}\frac{4x^2\,{\rm d}x}{y}
         \ =\ \frac{2\sqrt{2}}{\pi}
             \Vvevss{V_2(\infty)
               \mu(e_1)\mu(e_2)\mu(e_3)\mu(e_4)V_{-2}(0)}_{(e_2,e_3)}
             \nonumber\\
         &=& \frac{2\sqrt{2}}{\pi}\frac{e_1^2}
             {(e_3-e_2)^{\frac12}(e_4-e_1)^{\frac12}}
             \Vvevss{\mu(\infty)\mu(1)\mu(0)\mu(M(e_4))V_{-2}(M(0))
               V_2(M(\infty))}_{(0,1)}
             \nonumber\\
         &=& 2\sqrt{2}\frac{e_3^2}
             {(e_4-e_3)^{\frac12}(e_2-e_1)^{\frac12}}
             F_D^{(3)}({\textstyle\frac12,\frac12},-2,2,1;\xi,\eta,\varpi)\,,
  \eea
with the second inverse cross ratio $\eta=1/M(0)=
\frac{e_1(e_2-e_3)}{(e_1-e_2)e_3}$, and $\varpi=1/M(\infty)=
\frac{e_2-e_3}{e_2-e_1}$ the
inverse of the image of the double pole at infinity (which absorbs the zero
modes). The Lauricella $D$-type functions are generalized hypergeometric
functions in several variables, given as power series (where
$(a)_n=\Gamma(a+n)/\Gamma(a)$ is the Pochhammer symbol)
    \bea
    \lefteqn{F_D^{(n)}(a,b_1,b_2,\ldots,b_n,c;x_1,x_2,\ldots,x_n)=}\nonumber\\
    &&\sum_{m_1=0}^{\infty}\sum_{m_2=0}^{\infty}\ldots\sum_{m_n=0}^{\infty}
    \frac{(a)_{m_1+m_2+\ldots+m_n}(b_1)_{m_1}(b_2)_{m_2}\ldots(b_n)_{m_n}}
    {(c)_{m_1+m_2+\ldots+m_n}(1)_{m_1}(1)_{m_2}\ldots(1)_{m_n}}
    x_1^{m_1}x_2^{m_2}\ldots x_n^{m_n}\,,
  \eea
whenever $|x_1|,|x_2|,\ldots,|x_n|<1$. Its integral representation has the
form of a CFT screening integral,
$\frac{\Gamma(a)\Gamma(c-a)}{\Gamma(c)}
F_D^{(n)}(a,b_1,\ldots,b_n,c;x_1,\ldots,x_n) =
\int_0^1u^{a-1}(1-u)^{c-a-1}\prod_{i=1}^n(1-ux_i)^{-b_i}\,{\rm d}u$.
For $n=1$, it reduces to the ordinary Gauss hypergeometric
function ${}_2F_1(a,b_1;c;x_1)$, and for $n=2$, it is nothing else than
the Appell function $F_1(a;b_1,b_2;c;x_1,x_2)$. A great deal of information
on these functions may be found for example in the book \cite{Ext72} by
Exton. An important fact is that $F_D^{(n)}$ satisfies the following system
of partial differential equations of second order:
  \be
    \left[
    (1-x_j)\sum_{k=1}^nx_k\frac{\partial^2}{\partial x_k\partial x_j}
    + \left(c-(a+b_j+1)x_j\right)\frac{\partial}{\partial x_j}
    - b_j \sum_{{\scriptstyle k=1\atop\scriptstyle k\neq j}}^n
      x_k\frac{\partial}{\partial x_k}
    - ab_j\right]F = 0\,,
  \ee
where $j=1,\ldots,n$. Interestingly, this remains true even in the case
that massive hypermultiplets are present ($N_f>0$), while the Picard-Fuchs
equations now are of third order. However, the price paid is an artifically
enlarged number of variables. Furthermore, we easily can write down
differential equations of second and third order for each field in
the correlator which is proportional to $F_D^{(n)}$, depending on whether
the field is degenerate of level two, e.g.\ $\mu=\Psi_{1,2}$,
$V_{-1}=\Psi_{2,1}$, or three as $V_1=\Psi_{1,3}$ (where we consider the
$c=-2$ CFT as the degenerate model with $c=c_{2,1}$) according to \cite{BPZ83}.
We extensively exploit the special
properties of these functions in our forthcoming paper \cite{myself}.

Again, we may obtain the dual period by exchanging $e_2$ with $e_1$,
yielding
  \be
    a_D(u) = 2\sqrt{2}\frac{e_3^2}
             {(e_4-e_3)^{\frac12}(e_1-e_2)^{\frac12}}
             F_D^{(3)}({\textstyle\frac12,\frac12},-2,2,1;1-\xi,1-\eta,
             1-\varpi)\,.
  \ee
The two periods given above are by construction the
$a_{(\alpha)}$ and $a_{(\beta)}$ periods respectively. We will later also
need the period integrated between $e_2$ and $e_4$, which is
  \be\label{eq:adyon}
    a_{(2\alpha-\beta)}(u) = 2\sqrt{2}\frac{-e_2^2}
             {(e_4-e_3)^{\frac12}(e_1-e_2)^{\frac12}}
             F_D^{(3)}({\textstyle\frac12,\frac12,-2,2,1;
             1-\xi,\frac{\xi-1}{\eta-1},\frac{\xi-1}{\varpi-1}})
             \,.
  \ee
It is worth
noting that the dependency on three variables is superficial, since all
cross ratios are solely functions in the four branch points. Indeed,
we have $\xi=\varpi^2$, $\eta=-\varpi$. However, we
needed a fifth vertex operator in the CFT picture, located at zero, which
is the only singular point of the rational map $R(x)=\Lambda^4/A(x)^2$.
The inverse crossing ratios $\xi,\eta,\varpi$ have the nice property that they
tend to zero for $|u|\!\gg\!1$, e.g.\ $\xi\sim (\frac12\frac{\Lambda^2}{u})^2
+ O(u^{-4})$. Hence,
the overall asymptotics of $a(u)$ and $a_D(u)$ is entirely determined by
the prefactors, which are $a(u)\sim
\frac{2\sqrt{2}e_3^2}{\sqrt{e_4-e_3}\sqrt{e_2-e_1}}
\sim \sqrt{2u} + O(u^{-\frac12})$ and $a_D(u)\sim
\frac{\sqrt{2}e_3^2}{\pi\sqrt{e_4-e_3}\sqrt{e_1-e_2}}\log(\xi)\sim
\frac{i}{\pi}\sqrt{2u}\log(u) + O(u^{-\frac12}\log(u))$. Expanding
$a(u)$ as a power series in $1/u$ yields the familiar result
  \bea
    a(u) &=& \sqrt{2u}\left[1-\frac{1}{16}\frac{\Lambda^4}{u^2}
             -\frac{15}{1024}\frac{\Lambda^8}{u^4}
             -\frac{105}{16384}\frac{\Lambda^{12}}{u^6}
             -\frac{15015}{4194304}\frac{\Lambda^{16}}{u^8} + O(u^{-10})
             \right]\nonumber\\
         &=& \sqrt{2}\sqrt{u+\Lambda^2}\,{}_2F_1({\ts-\frac12,\frac12,1};
             \frac{2\Lambda^2}{u+\Lambda^2})\,.
  \eea

The strength of the CFT picture becomes apparent when asymptotic regions
of the moduli space are to be explored. Then, OPE and fusion rules provide
easy and suggestive tools. For example, the asymptotics of $a(u)$ and
$a_D(u)$ follow directly from the OPE of the field $\mu$ as discussed in
the preceeding section. The logarithmic partners of primary fields appear
precisely, if the contour of the screening charge integration gets pinched
between the two fields whose OPE is inserted.
Thus, the choice of contour together with the choice of internal
channels (due to the inserted OPEs) determines which term
of the OPE $\mu(z)\mu(0)\sim z^{1/4}(V_1(0) + \Lambda_1(0) - 2\log(z)V_1(0)
+\ldots)$ is picked. The three terms, which all have the same scaling
dimension $h=0$, correspond to the three possibilities of two branch
points flowing together. Either, they belong to different cuts such
that two cuts become one, or they belong to the same cut which becomes
a pole. The third case arises if they pinch a contour between them.
For example, when expanded in $\xi$, both periods,
$a(u)$ and $a_D(u)$ have asymptotics according to inserting the OPEs
$\mu(e_2)\mu(e_3)$ and $\mu(e_1)\mu(e_4)$. Keeping in mind (\ref{eq:vvev})
when inserting an OPE, we find with $e_{ij}=e_i-e_j$
  \bea
  a(u)&\sim&\left[e_{12}e_{13}e_{42}e_{43}\right]^{-1/4}
            \frac{e_1e_2}{e_3e_4}\left[e_{34}\Vvevss{V_2(\infty)
            \Lambda_1(e_3)V_1(e_4)V_{-2}(0)} + \ldots\right]
            \nonumber\\
      &\sim&\left[e_{12}e_{13}e_{42}e_{43}\right]^{-1/4}
            \frac{e_1e_2e_4}{e_3}\left[\Vvevss{V_2(\infty)
            \Lambda_1(e_4)V_{-2}(0)} + \ldots\right]
            \nonumber\\
      &\sim&\sqrt{2u} + \ldots\,,
  \eea
where the three-point functions evaluate trivially.
In a similar fashion, we obtain
  \bea
  a_D(u)&\sim&\frac{1}{i\pi}\left[e_{12}e_{13}e_{42}e_{43}\right]^{-1/4}
            \frac{e_1e_2}{e_3e_4}\left[e_{34}\Vvevss{V_2(\infty)
            \Lambda_1(e_3)\Lambda_1(e_4)V_{-2}(0)} + \ldots\right]
            \nonumber\\
      &\sim&\frac{1}{i\pi}
            \left[e_{12}e_{13}e_{42}e_{43}\right]^{-1/4}
            \frac{e_1e_2e_4}{e_3}\left[-2\log(e_4-e_3)\Vvevss{V_2(\infty)
            \Lambda_1(e_3)V_{-2}(0)} + \ldots\right]
            \nonumber\\
      &\sim&\frac{i}{\pi}\sqrt{2u}\,[\log(u) + 2\log(2) + \ldots]\,.
  \eea
Of course, other internal channels can be considered. In particular,
we may insert the OPE for $|e_1-e_3|\ll 1$ to get the behavior of the
periods for the case $u\longrightarrow\Lambda^2$. In fact, $a_D(u)$ and
$a(u)$ exchange their r\^ole since now the monopole becomes massless.
Put differently, duality in Seiberg-Witten models cooks down to
crossing symmetry in our $c=-2$ LCFT. The leading term can be read off from
$a_D(u)$ above (the OPE factors turn out to be the same upto a braiding
phase) to be proportional to $i(u-\Lambda^2)/\sqrt{2\Lambda^2}$. The relative
normalization of the logarithmic operator $\Lambda_1$ with respect to its
primary partner is fixed by the requirement that $a_D(u)$ is
the analytic continuation of $a(u)$ via crossing symmetry yielding the
factor of $(i\pi)^{-1}$.

There is one further BPS state which can become massless, since there is
one further zero of the discriminant
  \be
  \Delta(y^2(x))=(\det\bar{\partial}_{(j=\frac12)})^{8}=
  \left(\Vev{\prod_{i=1}^{2g+2}V_{1/2}(e_i)}_{c=1}\right)^{8}
  =\prod_{j<k}(e_j-e_k)^2\,,
  \ee
namely $e_2\longrightarrow e_4$. This is a dyonic state with charge
$(q,g)=(-2,1)$, meaning that both, the $\alpha$ cycles as well as the
$\beta$ cycle, get pinched in this limit.
It follows that both, $a(u)$
as well as $a_D(u)$, will receive logarithmic corrections when
$u\longrightarrow -\Lambda^2$, which is well known to be the case.

Within the CFT picture, higher gauge groups as well as additional,
possibly massive, flavours are treated on the same footing. Hence, we
obtain for the $SU(2)$ case with $N_f<4$ hypermultiplets, after
simple algebra in the numerator,
  \bea
    \lambda_{{\rm SW}} &=& \frac{1}{2\pi i}
    \frac{x\,{\rm d}x}{y\prod_{k=1}^{N_f}(x-m_k)}\left(
    4x\prod_{k=1}^{N_f}(x-m_k) - (x-\sqrt{u})(x+\sqrt{u})\sum_{k=1}^{N_f}
    \prod_{l\neq k}(x-m_l)\right)\nonumber\\
    &=&\frac{{\rm d}x}{2\pi i}\left(\frac{4x^2}{y}-\sum_{k=1}^{N_f}
    \frac{x(x-\sqrt{u})(x+\sqrt{u})}{y(x-m_k)}\right)\nonumber\\
    &=& \frac{{\rm d}x}{2\pi i}\left((4-N_f)\frac{x^2}{y} + N_f\frac{u}{y}
    -\sum_{k=1}^{N_f}m_k\left(\frac{x^2}{y(x-m_k)} - \frac{u}{y(x-m_k)}
    \right)\right)\,,
  \eea
such that we immediately can express the periods of the Seiberg-Witten form
in 4-point and 5-point functions. To this end we use $\frac{x^2}{y(x-m_k)} =
\frac{x+m_k}{y}+\frac{m_k^2}{y(x-m_k)}$ to rewrite the last term, and
obtain
  \bea\lefteqn{
    \oint\lambda_{{\rm SW}} = \frac{1}{2\pi i}\left(
    (4-N_f)\Vvevss{V_2(\infty)\mu(e_1)\mu(e_2)\mu(e_3)\mu(e_4)V_{-2}(0)}
    \vphantom{\sum_{k=1}^{N_f}}\right.}\nonumber\\
    &+&uN_f\Vvevss{\mu(e_1)\mu(e_2)\mu(e_3)\mu(e_4)}
    -\sum_{k=1}^{N_f}m_k
    \Vvevss{V_1(\infty)\mu(e_1)\mu(e_2)\mu(e_3)\mu(e_4)V_{-1}(-m_k)}
    \nonumber\\
    &+&\left.\sum_{k=1}^{N_f}m_k(u-m_k^2)
    \Vvevss{V_{-1}(\infty)\mu(e_1)\mu(e_2)\mu(e_3)\mu(e_4)V_1(m_k)}
    \right)\,.
  \eea
We recover hence the well know result that for all $m_k=0$ the scalar
modes have roughly the same form as in the $N_f=0$ case.
Including the charge balance
at infinity, this leads to the following expression ($x(\cdot) = 1/M(\cdot)$
denote the inverse crossing ratios)
  \bea
    \oint\lambda_{{\rm SW}} &=& \left(
      \frac{(4-N_f)e_3^2}{(e_4-e_3)^{\frac12}(e_2-e_1)^{\frac12}}
      F_D^{(3)}({\ts\frac12,\frac12,-2,2,1};x(e_4),x(0),x(\infty))
      \vphantom{\sum_{k=1}^{N_f}}\right.\\
    &+&\frac{uN_f}{(e_2-e_1)^{\frac12}(e_4-e_3)^{\frac12}}
      \,{}_2F_1({\ts\frac12,\frac12;1};x(e_4))\nonumber\\
    &-& \sum_{k=1}^{N_f}\frac{m_k(e_3+m_k)}
      {(e_2-e_1)^{\frac12}(e_4-e_3)^{\frac12}}
      F_D^{(3)}({\ts\frac12,\frac12,-1,1,1};
      x(e_4),x(-m_k),x(\infty))\nonumber\\
    &+& \left.\sum_{k=1}^{N_f}\frac{m_k(u-m_k^2)}
      {(e_2-e_1)^{\frac12}(e_4-e_3)^{\frac12}(e_3-m_k)}
      F_D^{(3)}({\ts\frac12,\frac12,1,-1,1};
      x(e_4),x(m_k),x(\infty))\right)\,.\nonumber
  \eea
Since the $F_D^{(3)}$ Lauricella functions have a negative integer
as one of the numerator parameters, they can be expanded as
polynomials in $F_1$ Appell functions, i.e.\ 5-point functions via
\be
    F_D^{(3)}(a;b,b',b'';c;x,y,z) =
    \sum_{m=0}^{\infty}\frac{(a)_m(b')_my^m}{(1)_m(c)_m}\,
    F_1(a+m;b,b'';c+m;x,z)\,,
  \ee
since this expansion truncates for $b'\in\BZ_-$.
Of course, we could have expressed this from the beginning
by only one correlation function proportional to
$F_D^{(2N_f+3)}$ of $2N_f+3$ variables, as indicated in (\ref{eq:periods}),
which is to be contrasted with the approach taken in \cite{Cappelli}.

As one further example, we obtain for $SU(3)$ without hypermultiplets,
where $R(Z) = \Lambda^6/(Z^3 - uZ + v)^2$ such that the resulting hyperelliptic
curve has six branch points $e_i$ and its metric $|\lambda_{{\rm SW}}|^2$
possesses three zeroes $z_j$, the solution
  \bea\lefteqn{
  \oint_{\gamma}\lambda_{{\rm SW}} = 2\Vvevs{V_{2}(\infty)\mu(e_1)\ldots
     \mu(e_6)V_{-1}(-\sqrt{u/3})V_{-1}(0)V_{-1}(\sqrt{u/3}}_{(\gamma)}}\\
  &=&\prod_{i=1}^3(\partial_{e_i}M(e_i))^{\frac14}
     \prod_{i=4}^6\left(\frac{\partial_{e_i}M(e_i)}{M(e_i)^2}\right)^{\frac14}
     \prod_{j=1}^3\left(\frac{\partial_{z_j}M(z_j)}{M(z_j)^2}\right)^{-\frac12}
     \lim_{z\rightarrow\infty}\left(
     \frac{z^2\partial_{z}M(z)}{M(z)^2}\right)\nonumber\\
  &\times& F_D^{(7)}
    ({\textstyle\frac12,\frac12,\frac12,\frac12,
    -1,-1,-1,2,1};
    x(e_4),x(e_5),x(e_6),
    x(0),x({\ts-\sqrt{u/3}}),x({\ts\sqrt{u/3}}),
    x(\infty))\,,\nonumber
  \eea
whith the last equality valid for $\gamma=\alpha_1\equiv C(e_2,e_3)$.
This Lauricella $D$-system for seven variables provides the complete
set of all periods. There exist more compact expressions in the
literature for this case, where the Appell function $F_4$ is involved
\cite{KlemmLerche}. However, presenting the solution in this way is more
transparent, if we view the moduli space of low-energy effective
field theory as created from string- or $M$-theory, e.g.\ as intersecting
$NS$-5 and $D$-4 branes. Then, the branch points $e_i$ and mass poles
$m_k$ are the directly given data -- they denote the endpoints of the
intersections. It remains to interpret the zeroes of the Seiberg-Witten form
within the brane picture, since they appear on equal footing with the
other singular points in our CFT approach. Moreover, this approach
suggests that BPS states from geodesic integration paths \cite{geodesics}
joining two
zeroes of $\lambda_{{\rm SW}}$ can be described in much the same way
as the more familiar BPS states connected to the periods. The zeroes
of $\lambda_{{\rm SW}}$ correspond to branching points in the fibration
of Calabi-Yau threefold compactifications of type II string theory, and
the corresponding BPS states are related to 2-branes ending on the
5-brane worldvolume $\BR^4\times\Sigma$.

Expressing the Seiberg-Witten periods in terms of
correlation functions reveals a further complication in exploring the
moduli space of low-energy effective field theories. These periods depend
only on the moduli $s_k$ and perhaps masses $m_l$. So, for the $SU(3)$
example above, the periods really depend only on two variables, $u,v$.
However, $\lambda_{{\rm SW}}$ in its factorized form naturally leads
to a 10-point function! The complete set of solutions of the
associated Lauricella $F_D^{(7)}$ system which covers all of $\BC^7$ is
actually quite large, and exceeds by far the set of periods obtainable from
simple paths enclosing two of the singular points (Pochhamer paths).
As is demonstrated in \cite{Ext72}, one needs in addition at least
so-called trefoil loops which are self-intersecting contours dividing the
set of singularities into three disjunct groups.

The reason behind all this enrichment is buried in the fact that we are
dealing with a Riemann surface together with an associated metric
$\lambda_{{\rm SW}}$.
A detailed analysis of all these features relies on a deeper knowledge of the
analytic properties of Lauricella functions and
will be carried out in our forthcoming paper \cite{myself}.

\bigskip\nn{\sc Acknowledgment:} I have enjoyed many fruitful discussions with
and comments from Ralph Blumenhagen, Wolfgang Eholzer, Matthias Gaberdiel,
Amihay Hanani, Andreas Honecker, Horst Kausch, Neil Lambert, Werner Nahm,
Gerard Watts, and Peter West. This research is funded by the
EU TMR network number FMRX-CT96-0012.

%
%
  
  \end{document}